\documentclass[12pt,a4paper]{article}
\usepackage{amsmath,pifont,amsfonts,amssymb,latexsym,multicol}
\usepackage{float,xspace,calc,theorem,ifthen}
\usepackage{fancyhdr,enumerate,psboxit,epsfig}
\usepackage{times,epic,amscd} 
\usepackage{version,xr}
\usepackage{aliascnt}
\usepackage{hyperref}
\usepackage{color}
\hypersetup{colorlinks}
\numberwithin{equation}{section}
\numberwithin{figure}{section}
\theoremstyle{change}
\newtheorem{theorem}{Theorem} [section]
\newtheorem {lemma}[theorem]{Lemma}
\newtheorem {prop}[theorem]{Proposition}

\newtheorem {corollary}[theorem]{Corollary}

{\theorembodyfont{\normalfont}\newtheorem {remark}[theorem]{Remark}}
{\theorembodyfont{\normalfont}\newtheorem {remarks}[theorem]{Remarks}}
{\theorembodyfont{\normalfont}\newtheorem {example}[theorem]{Example}}
\newcommand{\IF}{\ensuremath{\mathbb{F}}}
\newcommand{\IH}{\ensuremath{\mathbb{H}}}
\newcommand{\IN}{\ensuremath{\mathbb{N}}}
\newcommand{\IP}{\ensuremath{\mathbb{P}}}

\newcommand{\IR}{\ensuremath{\mathbb{R}}}

\newcommand{\IZ}{\ensuremath{\mathbb{Z}}}

\newcommand{\beq}{\begin{equation}}
\newcommand{\eeq}{\end{equation}}
\newcommand{\Leq}[1]{\label{#1}\end{equation}}
\newcommand{\beqn}{\begin{eqnarray}}
\newcommand{\eeqn}{\end{eqnarray}}
\newcommand{\beqno}{\begin{eqnarray*}}
\newcommand{\eeqno}{\end{eqnarray*}}
\newcommand{\qmbox}[1]{\quad\mbox{#1}\quad}
\newcommand {\eh}{{\textstyle \frac{1}{2}}}
\providecommand{\idty}{{\rm 1\mskip-4mu l}} 

\begin{document}

\title{Stable Degeneracies for Ising Models}

\author{Andreas Knauf\thanks{
Department of Mathematics, Friedrich-Alexander-University Er\-langen-Nuremberg, 
Cauerstr.\ 11, D-91058 Erlangen, Germany,  \texttt{knauf@math.fau.de}}
}
\date{August 12, 2015}
\maketitle

\begin{abstract}
We introduce and consider the notion of {\em stable degeneracies} of 
translation invariant energy functions for finite Ising models.  By this term we 
mean the lack of injectivity that cannot be lifted by changing the interaction.
 
We show that besides the symmetry-induced degeneracies, related to
spin flip, translation and reflection, there exist additional stable degeneracies, due
to more subtle symmetries. One such symmetry is the one of the
Singer group of a finite projective plane. Others are described by combinatorial relations
akin to trace identities.

Our results resemble traits of the
length spectrum for closed geodesics on a Riemannian surface of constant
negative curvature.  There stable degeneracy is defined w.r.t.\ Teichm\"uller
space as parameter space.
\end{abstract}

\tableofcontents
%
\section{Introduction}
%
The energy degeneracies of Ising models are of physical and 
mathematical relevance. 
Like in quantum mechanics, they can result from symmetries of the model. 
If the spins of these models are enumerated by an abelian group $F$,
one such symmetry is the translation invariance of the interaction.
We assume here that $F$ is finite. Then an {\em Ising model} is defined by the 
choice of the real coefficients $\tilde{j}_{f,g}$ and $\tilde{h}_{f}$ in the 
{\em energy function}\,\footnote{we omit the conventional negative sign here.}
\[H:\{-1,1\}^{F}\to \IR\qmbox{,}
\textstyle H(\sigma)
= \sum_{f\neq g\in F}\tilde{j}_{f,g}\sigma_f\sigma_g\ +\
\sum_{f\in F}\tilde{h}_{f}\sigma_f.\]
This absence of $n$--body interactions between $n\ge3$ spins
is physically realistic.
{\em Translation invariance} means $\tilde{j}_{f,g}\equiv j_{g-f}$
and $\tilde{h}_f\equiv h$. This, too is a realistic assumption for many
physical systems\,\footnote{although it excludes models with frustrated or
random interactions, see, {\em e.g.} the study of frustrated ground state degeneracy by 
Loebl and Vondr\'ak \cite{LV}.}.

For an Ising model in $d$ spatial
dimensions a typical choice for $F$ is $(\IZ/N\IZ)^d$.

\begin{remark}[Nearest neighbour interactions] \ \label{rem:n:n}
An additional feature that leads to large degeneracies
is that only {\em nearest neighbours} 
interact\,\footnote{that is, $\tilde{j}_{f,g}=0$ if $\|f-g\|_1>1$
for the norm induced by the Lee norm $\|a\|:=\min(a,N-a)$ ($a\in\IZ/N\IZ\cong
\{0,1,\ldots,N-1\}$) on the factors of $F$.}. 
There are $2^{|F|}$ spin configurations,
but in the $d$--dimensional case there are only $d|F|$ edges
in the nearest neighbour graph on the vertex set $F=(\IZ/N\IZ)^d$.

So for {\em isotropic} such interactions (meaning $j_\ell=j$ for
$\|\ell\|_1=1$ and $j_\ell=0$ otherwise) and
$h=0$ there at most $d|F|$ energy values, all multiples of $j$.
Their mean degeneracy is thus greater or equal to 
$2^{|F|}/(d|F|)$, growing exponentially in the thermodynamic limit
$N\to\infty$. A similar estimate applies for $h\neq 0$.

For the 'classical' $d=2$ dimensional Ising model
($F=(\IZ/N\IZ)^2$, 
$H(\sigma)= - \sum_{f\in F} \sigma_f(\sigma_{f+(1,0)}+\sigma_{f+(0,1)})$)
the degeneracies of $H$ were studied in Beale \cite{Be}, based on the
celebrated Kaufman-Onsager solution \cite{Ka}.

However, although models with nearest neighbour interactions are 
a bit easier to analyze than models with general 
two-body interactions, they are unrealistic. Physical interactions decay as 
the distance of the spins increases, but there is no reason why they should be  
of finite range.
\hfill $\Diamond$
\end{remark}
Here we consider the degeneracies of translation invariant
Ising spin models, but we allow for two-body interactions between
all pairs of spins. This realistic assumption
leads to an enormous decrease of degeneracies, independent of the
rate of spatial decay of these  interactions.

We are particularly interested in {\em stable} degeneracies, that is
degeneracies that cannot be lifted by changing the translation invariant interaction.

Although typically in the literature one considers $d$--dimensional
Ising models with groups $F=(\IZ/N\IZ)^d$, we just assume that $F$ is
finite abelian. 
Then the {\em configuration space} is the multiplicative group 
$G\equiv G_F:= \{-1,1\}^{F}$.

The energy function of a translation invariant Ising model 
has $J\equiv J_F:=\IR^{F}$ as parameter space\,\footnote{but compare with
Lemma \ref{lem:span:even} below.}
(assuming for now vanishing of $h$) and is of the form 
\beq
H\equiv H_F:G_F\times J_F\to \IR\qmbox{,} 
H_F(\sigma,j)=\sum_{f\in F} j_f
\sum_{\ell\,\in\,F}\ \sigma_\ell\sigma_{\ell+f}.
\Leq{energy}
For {\em interaction} $j\in J$ the 
{\em $j$--degeneracy} of $\sigma\in G$ is defined as 
\beq
D(\sigma,j) := \big|\big\{\tau\in G\mid H(\tau,j)= H(\sigma,j)\big\}\big|.
\Leq{j:degeneracy} 
Its {\em stable degeneracy} is a lower bound for $D(\sigma,j)$. 
We define it by 
\beq
D_{\rm stab}(\sigma):= \big|\big\{ \tau\in G\mid \forall\,j\in J: H(\tau,j)
= H(\sigma,j)\big\}\big|
= \big| A_F^{-1}\big(A_F(\sigma)\big) \big| 
\Leq{Dstab}
with the {\em correlation map}
\beq
A\equiv A_F:G_F\to \IZ^{F} \qmbox{,}
A_F(\sigma)_f =\sum_{\ell\,\in\, F}\ \sigma_\ell\sigma_{\ell+f},
\Leq{A:N}
since we have $H(\sigma,j)=\langle j,A(\sigma)\rangle$. 
This shows that for generic interactions $j$ we have $D(\sigma,j)=D_{\rm stab}(\sigma)$, 
see Remark \ref{rem:generic}.

\bigskip
\noindent
{\bf Results.}
In Section \ref{sect:2} we consider a lower bound $D_{\rm sym}$
of $D_{\rm stab}$ that is given by the 'obvious' symmetries of the 
energy function (that is, spin flip, translations,  and reflection) and thus
is of order ${\cal O}(|F|)$. Actually in the limit of many spins
typically there are $4|F|$ configurations related by these
symmetries (Prop.\ \ref{prop:mean:Dsym}).
Product configurations are a simple example of configurations with 
additional degeneracies (Proposition \ref{prop:prod}).

In Section \ref{sect:3} we address the question by how much stable
degeneracy can deviate from $D_{\rm sym}$. Empirical data 
are compatible with the supposition that typically there are no additional
degeneracies unrelated to the above-mentioned symmetries 
(Remark \ref{rem:empirical}). However, in Section \ref{sect:4} 
we present an infinite
family of configurations $\sigma$ (related to finite projective spaces) 
where $D_{\rm stab}(\sigma)$ essentially 
equals $|F|^2\gg 4|F|$. Alternatively Proposition \ref{prop:subs} of Section \ref{sect:5}
uses substitution techniques to generate
spin chain configurations with large stable degeneracy.
Finally, in Section \ref{sect:6}, it is shown that for up to four blocks
of equal adjacent spins nontrivial stable degeneracy does not occur 
(Proposition \ref{prop:four:blocks}), and that under an injectivity condition the same is
true for an arbitrary number of blocks (Proposition \ref{prop:reconstruction}).
\hfill$\Diamond$
\begin{remark} [Stable degeneracies for closed geodesics] 
\ \ In \cite{Ra} Randol showed that the length spectrum of the closed geodesics
on a Riemann surface of  constant negative curvature is of unbounded
multiplicity, independent on the point on Teichm\"uller space 
fixing the Riemannian metric.

This was based on a result by Horowitz \cite{Ho}, who considered
the free group generated by two elements of ${\rm SL}(2,\IR)$. 
He showed that for any $N\in \IN$ there exist $N$ non-conjugate words of 
letters in this generator that encode closed geodesics 
having the same length. These words are related by substitutions.

In Section \ref{sect:5} we adapt that method to the energy spectrum
of the translation invariant Ising model. 
We show the existence of sequences of spin configurations,
whose quotient of stable and symmetry induced degeneracy is unbounded.

It should be noted, however that stable degeneracy 
of closed geodesics and of spin configurations are very different
phenomena. Whereas the number of interaction parameters  is proportional
to the number $N$ of spins and thus diverges in the thermodynamic limit $N\to\infty$, 
the number of parameters for the geodesic problem is bounded
by the number of generators of the discrete subgroup $\Gamma$ of
${\rm SL}(2,\IR)$ that determines the surface $\Gamma\setminus \IH$.

Thus, in a way, the occurrence of non-trivial stable degeneracies for Ising models
is even more astonishing than the one for closed geodesics.
\hfill $\Diamond$
\end{remark}
{\bf Acknowledgement.}
The article originated from discussions with Catherine Meusburger (FAU Erlangen-N\"urnberg)
about moduli spaces of Riemannian surfaces and length spectra of geodesics.
Many thanks to her!
%
\section{Properties of the Correlation Map $A_F$} \label{sect:2}
%
In order to understand stable degeneracy, we must study the level sets
of $A_F$. There are three obvious types of symmetries, leaving $A$ invariant:
\begin{itemize}
\item 
The {\em spin flip} $G\to G,\ \sigma\mapsto -\sigma$ induces an action
$S:\{-1,1\}\times G\to G$.
\item 
$F$ acts on itself by {\em translations}. This induces the action 
\[T:F\times G\to G,\ \big(T_t(\sigma)\big)_f=\sigma_{f+t}
\qmbox{, and} A\circ T_t=A.\]
\item 
Finally, the automorphism group ${\rm Aut}(F)$ of $F$ acts on $G$.
Whereas this does not in general leave $A$ invariant, this is the case for
the {\em reflection} $F\to F$, $f\mapsto -f$. 
This gives rise to an action $R$ of $\{-1,1\}$
on $G$.
\end{itemize}
Altogether we obtain an action
\beq
\Phi :=(S,T,R):{\mathcal S}\times G\to G
\qmbox{,}
 \big(\Phi_{(s,t,r)}(\sigma) \big)_f
 =s \sigma_{rf +t}
\Leq{Phi}
of the group ($\rtimes$ denoting semidirect product)
\[{\mathcal S}\equiv {\mathcal S}_F := \{-1,1\}\times \left(F\rtimes \{-1,1\}\right).\]
This action is faithful iff the group exponent of $F$ is at least 3, that is,
unless $F\cong (\IZ/2\IZ)^d$ for some $d\in\IN_0$.
The $\Phi$--orbits have cardinalities dividing the group order $|{\mathcal S}_F|=4|F|$:
\beq
D_{\rm sym}(\sigma):=|\Phi({\mathcal S},\sigma)|
= |{\mathcal S}_F|/ |{\mathcal S}_F(\sigma)|
\qquad (\sigma\in G) ,
\Leq{D:sym}
with the stabilizer group of $\sigma$
\[{\mathcal S}_F(\sigma):= 
\{(s,t,r)\in{\mathcal S}_F\mid \Phi_{(s,t,r)}(\sigma)=\sigma\}.\] 
Since $A\circ \Phi_s=A$ \ $(s\in {\mathcal S})$,
this leads to a lower bound for stable degeneracy:
\beq 
D_{\rm stab}\ge D_{\rm sym}.
\Leq{DD0}
As the examples $\sigma=\pm\idty_F$ show, there are $\Phi$--orbits of size two.
However, we show now that for groups $F$ of large order typically 
$D_{\rm sym}(\sigma) = |{\mathcal S}_F|$. More precisely, convergence 
in the mean occurs as the group order of $|F|$
goes to infinity, unless reflection $R$ acts trivially:
\begin{prop} [Average symmetry-induced degeneracy] \label{prop:mean:Dsym}
For the family of finite abelian groups $F$ of group exponents $\ge3$,
uniformly in the group order
\[\lim_{|F|\to\infty}
\frac{|G_F|^{-1}\sum_{\sigma\in G_F}D_{\rm sym}(\sigma)}{|{\mathcal S}_F|} = 1.\]
\end{prop}
\textbf{Proof:} 
The upper bound 
$\sum_{\sigma\in G_F}D_{\rm sym}(\sigma) \le {|G_F|\, |{\mathcal S}_F|}$ being obvious,
we need a lower bound. As 
$D_{\rm sym}(\sigma)= |{\mathcal S}_F|/ |{\mathcal S}_F(\sigma)|$,
we are to show that typically the stabilizer group 
${\mathcal S}_F(\sigma)$ of $\sigma\in G_F$ is trivial. 
\begin{enumerate}
\item 
For all {\em spin flips} $g=(-1,t,r)\in{\mathcal S}_F$ a necessary condition for 
$\Phi_{g}(\sigma)=\sigma$ is that
\[|\{f\in F\mid \sigma_f=1\}|=|F|/2.\] 
But by Stirling's formula, this can only be true for a subset of $G_F$
which is of order ${\mathcal O}\big(|F|^{-1/2}|G_F|\big)=o(|G_F|)$.
\item 
A {\em translation} ($g=(1,t,1)\in{\mathcal S}_F$ 
with $t\in F\setminus \{0\}$) spans a non-trivial subgroup $U$ of $F$, and
the set of $g$--invariant configurations is isomorphic to $G_{F/U}$. Thus it
is of order $|G_{F/U}|={\mathcal O}(2^{|F|/2})$, and 
\[\big|\{\sigma\in G_F\mid \exists t\in F\setminus \{0\}: 
\Phi_{(1,t,1)}(\sigma)=\sigma\}\big|
={\mathcal O}\big((|F|-1)2^{|F|/2}\big)=o(|G_F|).\]
\item 
By assumption $F\neq (\IZ/2\IZ)^d$. 
For a {\em reflection} ($g=(1,t,-1)\in{\mathcal S}_F$ with $t\in F$)
the fixed point set $F_g:=\{f\in F\mid 2f=t\}$ of the action of $g$ on $F$
has thus cardinality $|F_g|\le |F|/2$. 
So the set of $g$--invariant configurations is of order
${\mathcal O}\big(2^{|F_g|+|F\setminus F_g|/2}\big)
={\mathcal O}\big(2^{3|F|/4}\big)$.
Therefore
\[\big|\{\sigma\in G_F\mid \exists t\in F: \Phi_{(1,t,-1)}(\sigma)=\sigma\}\big|
={\mathcal O}\big(|F|2^{3|F|/4}\big)=o(|G_F|).\hfill \tag*{$\Box$}\]
\end{enumerate}
\begin{remark}[Group exponent]
The condition ${\rm Exp}(F)\ge3$ in Prop\ \ref{prop:mean:Dsym}
is necessary, as otherwise
$F$ is isomorphic to $(\IZ/2\IZ)^d$ for some $d\in\IN_0$. Then one has 
$\Phi_{(s,t,-1)}= \Phi_{(s,t,1)} \ \big((s,t)\in S\times F\big)$, so that 
$D_{\rm sym}(\sigma) \le 2|F|=\eh |{\mathcal S}_F|$.
\hfill $\Diamond$
\end{remark}
\begin{lemma} 
For all finite abelian groups $F$, the {\bf  mean stable degeneracy} 
\[{\rm MSD}(F) := \frac{|G_F|}{ |A_F(G_F)|} \]
is smaller than the {\bf average stable degeneracy}: 
\[  {\rm MSD}(F) \le
|G_F|^{-1}\sum_{\sigma\in G_F}D_{\rm stab}(\sigma). \]
\end{lemma}
\textbf{Proof:} 
This is an application of the Cauchy-Schwarz inequality:
\[{\rm MSD}(F)= \frac{\sum_{x\in A(G)}|A^{-1}(x)|}{\sum_{x\in A(G)}1}
\le \frac{\sum_{x\in A(G)}|A^{-1}(x)|^2}{\sum_{x\in A(G)}|A^{-1}(x)|}
= |G_F|^{-1}\sum_{\sigma\in G_F}D_{\rm stab}(\sigma),\]
since 
$\big(\sum_{x\in A(G)}|A^{-1}(x)|\big)^2 \le 
\big(\sum_{x\in A(G)}1\big)\big(\sum_{x\in A(G)}|A^{-1}(x)|^2\big)$.
\hfill $\Box$\\[2mm]
\begin{remark}[Averages of stable degeneracy] 
By \eqref{DD0}, the average stable degeneracy meets the estimate
\[\liminf_{|F|\to\infty} 
\frac{|G_F|^{-1}\sum_{\sigma\in G_F}D_{\rm stab}(\sigma)}{|{\mathcal S}_F|}\ge 1.\]
We show in Section \ref{sect:3} that the quotient
$F\to[1,\infty)$, $\sigma\mapsto D_{\rm stab}(\sigma)/D_{\rm sym}(\sigma)$ 
is not bounded as the group order $|F|$ goes to infinity.
Nevertheless we {\bf conjecture} that 
for the family of finite abelian groups $F$ of group exponents $\ge3$
\beq
\lim_{|F|\to\infty}{\rm MSD}(F)/ |{\mathcal S}_F| = 1\qmbox{and}
\lim_{|F|\to\infty} 
\frac{|G_F|^{-1}\sum_{\sigma\in G_F}D_{\rm stab}(\sigma)}{|{\mathcal S}_F|} =1,
\Leq{conjecture}
that is, that typically stable degeneracy does not exceed
symmetry-induced degeneracy. 
\hfill $\Diamond$
\end{remark}
\begin{lemma} [Parameter space] \label{lem:span:even}
The number of parameters of the energy function \eqref{energy} is smaller than $F$, and
equals\, $\dim(\IR^F_{\rm ev})$ with
\[{\rm span}_\IR\big(A_F(G_F)\big)=
\IR^F_{\rm ev} := \{h\in \IR^F \mid \forall f\in F:h_{-f}=h_f\}.\]
\end{lemma}
\textbf{Proof:}
As $A_F(\sigma)_{-m}=A_F(\sigma)_{m}$, 
${\rm span}_\IR(A_F(G_F))\subseteq \IR^F_{\rm ev}$. Conversely
we set
\[\sigma^{(f)}\in G_F\qmbox{,}\sigma^{(f)}_\ell
:=\left\{
\begin{array}{cl}
-1  & ,\ \ell=0\mbox{ or }\ell=f     \\
 1 &   ,\ \mbox{else} 
\end{array}
\right.\qquad(f\in F).\]
Then, with the characteristic function $\idty_S$ of a subset $S\subseteq F$, 
\beq
A_F(\idty) = |F|\ \idty_{F}\qmbox{,} 
A_F(\idty)-A_F(\sigma^{(0)}) =  4\ \idty_{F\setminus \{0\}},
\Leq{span1}
and for $f\in F\setminus\{0\}$
\[A_F(\sigma^{(f)})_m = \left\{
\begin{array}{ll}
|F|& ,\  m=0\mbox{ or }(2f=0\mbox{ and }m=f)\\
|F|-4& ,\ 2f\neq 0\mbox{ and }m\in\{-f,f\}\\
|F|-8& ,\  \mbox{else }\\
\end{array}
\right. .\]
Thus for $f\in F\setminus\{0\}$
\[\big[A_F(\sigma^{(f)})-A_F(\sigma^{(0)})\big]_m = 
\left\{
\begin{array}{ll}
4& ,\  2f=0\mbox{ and }m=f\\
0& ,\  m=0\mbox{ or }(2f\neq0\mbox{ and }m\in\{-f,f\})\\
-4& ,\  \mbox{else }
\end{array}
\right. ,\]
so that with \eqref{span1}
\beq
A_F(\sigma^{(f)})-2A_F(\sigma^{(0)}) + A_F(\idty) 
= 4(\idty_{\{f\}}+\idty_{\{-f\}}).
\Leq{span2}
By \eqref{span1} and \eqref{span2} the functions $A_F(\idty)$ and 
$A_F(\sigma^{(f)})$ ($f\in F$) span $\IR^F_{\rm ev}$.
\hfill $\Box$\\[2mm]
So with the orthogonal projection 
$J=\IR^F\to\IR^F_{\rm ev},\  j\mapsto j_{\rm ev}$
we have 
\[H(\sigma,j)=H(\sigma,j_{\rm ev})\qquad\big((\sigma,j)\in G\times J\big),\]
and we can substitute $\IR^F_{\rm ev}$ for $\IR^F$ in the definition 
\eqref{Dstab} of stable degeneracy.

According to the fundamental theorem of finite abelian groups, for 
(not necessarily distinct) primes $p_i$
\[F\ \cong\ {\textstyle \bigoplus_i }\ \IZ/p_{i}^{n_i}\IZ\]
(and we henceforth omit the isomorphism). 
So it is natural to consider for an arbitrary representation 
\beq
F= {\textstyle\bigoplus_{j=1}^d} F_j
\Leq{direct:sum}
of $F$ as a direct sum of finite
non-trivial abelian groups $F_j$ the behavior of the 
stable degeneracy $D$ and of its lower bound $D_{\rm sym}$ under such a 
decomposition.
The subgroup lattice for a decomposition \eqref{direct:sum} of a group $F$ 
into its {\em $p$--groups} (and more generally if the group orders 
$|F_j|$ are coprime) is multiplicative.
It contains more subgroups (and is quite complicated) 
for a decomposition \eqref{direct:sum}
{\em of}  a $p$--group $F$ into its factors $\IZ/p^{n_i}\IZ$, see C\u{a}lug\u{a}reanu \cite{Ca}.

Additionally the correlation $A_F$ of the direct sum \eqref{direct:sum} 
is in general {\em not} invariant under reflections $f_j\mapsto -f_j$
of a single group $F_j$.
However, this is the case 
for $A_F(\sigma)$ if $\sigma$ is the {\em product configuration}
\beq
\sigma\in G_{F}\qmbox{,}\sigma:=\otimes_{j=1}^d \sigma^{(j)}
\qquad \mbox{ of } \sigma^{(j)}\in G_{F_j}\quad (j=1,\ldots,d).
\Leq{product:conf}
\begin{prop}[Product configurations]  \label{prop:prod}
The correlation of the product configuration $\sigma$ is multiplicative, i.e.\
$A_F(\sigma)=\otimes_{j=1}^d A_{F_j}(\sigma^{(j)})$ and
\begin{enumerate}
\item 
$4^{1-d} \prod_{j=1}^d D_{\rm sym}(\sigma^{(j)})\le D_{\rm sym}(\sigma) \le
\prod_{j=1}^d D_{\rm sym}(\sigma^{(j)})$,
and for all $d$ both inequalities cannot be improved.
\item 
The stable degeneracy obeys the inequality
$D_{\rm stab}(\sigma)\ge 2^{1-d}\prod_{j=1}^d D_{\rm stab}(\sigma^{(j)})$.
\end{enumerate}
\end{prop}
\textbf{Proof:} 
For all $f=(f_1,\ldots, f_d)\in F$
\[A_F(\sigma)_f=\sum_{\ell\in F}\sigma_\ell \sigma_{\ell+f}
=\sum_{\ell_1\in F_1,\ldots,\ell_d\in F_d}\prod_{j=1}^d  
\left(\sigma^{(j)}_{\ell_j}\sigma^{(j)}_{\ell_j+f_j}\right)
=\prod_{j=1}^d A_{F_j}(\sigma^{(j)})_{f_j}.\]
1.) 
The cardinalities are given by
$D_{\rm sym}(\sigma) = |{\mathcal S}_F|/ |{\mathcal S}_F(\sigma)|$ and
$D_{\rm sym}(\sigma^{(j)})
= |{\mathcal S}_{F_j}|/ |{\mathcal S}_{F_j}(\sigma^{(j)})|$,
with $|{\mathcal S}_F|=4^{1-d}\prod_{j=1}^d  |{\mathcal S}_{F_j}|$.
So we have to show the inequalities
$ |{\mathcal S}_F(\sigma)| \le \prod_{j=1}^d |{\mathcal S}_{F_j}(\sigma^{(j)})|
\le  4^{d-1}|{\mathcal S}_F(\sigma)|$  for the stabilizer groups.

For the lower bound, we observe that for 
$(s,t,r)\in {\mathcal S}_F(\sigma)$ the quotient functions
$F_j\to \{-1,1\}$, $f_j\mapsto\sigma^{(j)}_{r f_j + t_j}/\sigma^{(j)}_{f_j}$ 
are constant.

The upper bound for $|{\mathcal S}_F(\sigma)|$ follows from the observation that 
as a subgroup of ${\mathcal S}_F= \{-1,1\}\times F\rtimes \{-1,1\}$
it is isomorphic to $S\times F' \rtimes R$, with $S$ and $R$ trivial or equal to 
$\{-1,1\}$, and $F'$ a subgroup of $F$. Similar decompositions 
with subgroups $F'_j$ of $F_j$ exist for ${\mathcal S}_{F_j}(\sigma^{(j)})$, and
$F'\cong \oplus_{j=1}^d F'_j$.

The left inequality in 1.) is saturated for $\sigma^{(j)}$ with trivial stabilizer groups.
The right inequality in 1.) is saturated for 
$\sigma^{(j)}=(1,-1)\in G_{F_j}$, $F_j= \IZ/2\IZ$.\\[2mm]
2.) 
Consider $\tau^{(j)}\in A_{F_j}^{-1}\big(A_{F_j}(\sigma^{(j)})\big)\ (j=1,\ldots,d)$.
Then by the first statement of the proposition
$\tau\in G_{F}$,\ $\tau_f:=\otimes_{j=1}^d \tau^{(j)}_{f_j}$ is in 
$A_{F}^{-1}\big(A_{F}(\sigma)\big)$. Conversely, given such a $\tau$,
the $\tau^{(j)}$ can be reconstructed only up to a factor $s^{(j)}\in\{-1,1\}$, 
with $\prod_{j=1}^d s^{(j)}=1$.
So the product map is $2^{d-1}$ to one.
There may be additional elements in $A_{F}^{-1}\big(A_{F}(\sigma)\big)$,
not of product form. This leads to a lower bound for $D_{\rm stab}(\sigma)$.
\hfill $\Box$
\begin{remark} [Periodic configurations] 
For a homomorphism $\pi: F\to F'$ of finite abelian groups, 
the pull-back $\pi^*:G_{F'}\to G_F$ relates the correlation maps via
\beq
A_F\circ\pi^* = |F|/|F'|\ \pi^* \circ A_{F'}=|U|\ \pi^*\circ A_{F'},
\Leq{AF:AFp}
with the subgroup $U:={\rm ker}(\pi)$ of $F$. Of course we can also
reverse this, setting $F':=F/U$ for a subgroup $U$  of $F$.
The lifted configurations $\pi^*\tau\in G_F$ are $U$--periodic.
More precisely, the group homomorphism $\pi_{\mathcal S}: {\mathcal S}_F\to {\mathcal S}_{F'}$,
$(s,t,r)\mapsto (s,\pi(t),r)$ relates the actions \eqref{Phi}: 
$\pi^*\circ \Phi_{\pi_{\mathcal S}(g)}= \Phi_g\circ \pi^* 
\ (g\in {\mathcal S}_F)$.
Thus 
\[D_{\rm sym}(\sigma)= D_{\rm sym}(\tau)\qmbox{and}D_{\rm stab}(\sigma)\ge D_{\rm stab}(\tau) \qmbox{for}\sigma:=\pi^*\tau.\] 
Moreover, $A_F(\sigma)$ is $U$--periodic if and only if $\sigma\in G_F$ is
$U$--periodic. This follows from \eqref{AF:AFp}, and conversely since for
the $U$--periodic correlations $A_F(\sigma)$ we have 
$A_F(\sigma)_u=A_F(\sigma)_0=|F|$ $\ (u\in U)$. 
That equality implies $\sigma_{k+u}=\sigma_{k}$ $\ (k\in F,\, u\in U)$.
\hfill $\Diamond$
\end{remark}
\begin{remark} [Positive Fourier transform]  
We denote unitary {\em Fourier transform} by
\[{\mathcal F}\equiv{\mathcal F}_F:\ell^2(F)\to \ell^2(F^*) \qmbox{,} 
({\mathcal F}g)_f=
|F|^{-1/2}\sum_{m\in F} g_m \chi_m^{(f)},\]
with the characters $\chi^{(f)}:F\to S^1$, using that the dual group $F^*$
of $F$ is isomorphic to $F$.
$\sigma\in G_F$ and  $A_F(\sigma)$ 
can be considered as real-valued functions on $F$, with
$A_F(\sigma)=\sigma* I(\sigma)$ for $(Ig)_f:=g_{-f}$ and 
$(g*h)_k :=\sum_{f\in F}g_f h_{k-f}$ convolution. 
Its Fourier transform is non-negative,
\[{\mathcal F}\big(A_F(\sigma)\big) =|F|^{1/2}\ |{\mathcal F}(\sigma)|^2\ge 0\qquad
\big(\sigma\in G_F\big),\]
since 
$|F|^{-1/2}{\mathcal F}(A_F(\sigma))={\mathcal F}(\sigma){\mathcal F}(I(\sigma))
={\mathcal F}(\sigma)I({\mathcal F}(\sigma))
= |{\mathcal F}(\sigma)|^2$.
\hfill $\Diamond$
\end{remark}
The (left) action $\Psi: {\rm Aut}(F)\times F\to F$ gives rise to the
actions on $G_F$ and on $ \IR^F_{\rm ev}$
\[ \Psi^{(G)}: {\rm Aut}(F)\times G_F\to G_F\qmbox{,} 
\Psi^{(G)}_g(\sigma)_f = \sigma_{ \Psi_{g^{-1}} (f)} \]
\[ \Psi^{({\rm ev})}: {\rm Aut}(F)\times \IR^F_{\rm ev}\to \IR^F_{\rm ev}
\qmbox{,} 
\Psi^{({\rm ev})}_g(h)_f = h_{ \Psi_{g^{-1}} (f)}.\]
Together with the  $\Phi$--action defined in \eqref{Phi}, we obtain an
action of the semidirect product ${\mathcal S}_F\rtimes{\rm Aut}(F)$, with
\beq
\Psi^{(G)}_g\circ \Phi_{(s,t,r)}= \Phi_{(s,\Psi_{g}(t),r)}\circ\Psi^{(G)}_g. 
\qquad \big((s,t,r)\in {\mathcal S},\, g\in{\rm Aut}(F)\big);
\Leq{eq:joint}
in particular $\Psi^{(G)}_g$ acts on the $\Phi$--orbits.

By Lemma \ref{lem:span:even} the image of 
the correlation map $A_F$ spans $\IR^F_{\rm ev}$. 
Although $A_F$ is not invariant with respect to $\Psi^{(G)}$, 
it has the following simple properties.
\begin{prop}[Image of the correlation map]  \label{prop:image}
$\bullet$
$A_F(G_F)\subseteq 
C_F$ for
\[ C_F:= \big\{ |F|-4k \mid k=0,\ldots,\lfloor |F|/2\rfloor \big\}^{F} \cap \IR^F_{\rm ev},\] 
with $|F|\, \idty_F\, \in A_F(G_F)$. \\
$\bullet$
Its image is in general not a convex\,\footnote{We call a subset 
$S\subseteq C_F$ of the discrete cube $C_F$ {\em convex}, if for $s_0,s_1\in S$
and $s_t:=(1-t)s_0+ts_1\in C_F$ for some $t\in (0,1)$ we have $c_t\in S$, too.}
 subset of $C_F$.\\
$\bullet$
$A_F(G_F)$ is invariant under the action $\Psi^{({\rm ev})}$ of 
${\rm Aut}(F)$ on $\IR^F_{\rm ev}$, and 
\beq
A_F\circ\Psi^{(G)}_g=\Psi^{({\rm ev})}_g \circ A_F \qquad 
\big(g\in {\rm Aut}(F)\big).
\Leq{equivar}
\end{prop}
\noindent
\textbf{Proof:}
$\bullet$
$A_F(\sigma)=|F|\; \idty_F$ iff $\sigma=\pm\idty_F$.
For all $k,r\in F$ the spin umklapp
$\sigma_r\mapsto -\sigma_r$ of the $r$--th spin, keeping the other spins fixed,
changes exactly two terms in the sum $A_F(\sigma)_k$,
by $\pm2$. Since $A_F(-\sigma)=A_F(\sigma)$, we can restrict ourselves to
$\sigma\in G_F$ with $|\{k\in F\mid \sigma_k=-1\}|\le \lfloor |F|/2\rfloor$.\\
$\bullet$
Convexity fails in the example of $F:=\IZ/4\IZ\equiv \{0,1,2,3\}$, since
\[A_F\big((1,1,1,1)\big) = (4,4,4,4)\qmbox{and}
A_F\big((1,-1,1,-1)\big) = (4,-4,4,-4),\] 
but $(4,0,4,0)\not\in A_F(G_F)$.\\
$\bullet$
The $A_F$--equivariance \eqref{equivar} of the ${\rm Aut}(F)$--actions 
is immediate and implies
$\Psi^{({\rm ev})}_g \big(A_F(G_F) \big) = A_F\big(\Psi^{(G)}_g(G_F)\big) 
= A_F(G_F)$. 
\hfill $\Box$\\[2mm]
Lack of convexity makes it hard to find a good lower bound on the
size $|A_F(G_F)|$ of the image. Such a bound would be needed to prove 
conjecture \eqref{conjecture}.
\begin{example}[Nearest neighbour interaction]  
We consider $F=\IZ/N \IZ$.
If $j(1)=1$ but $j(d)=0$ for $d\in F\setminus\{1\}$, then 
the energy values are $H(G_F,j)=\{N-4k\mid k=0,\ldots,\lfloor N/2\rfloor\}$,
and for $h\in H(G_F,j)$ the degeneracy equals $|H(\cdot,j)^{-1}(h)| = 2{N\choose (h+N)/2}$.
Thus the {\em mean degeneracy} of the energy levels is asymptotic to $2^{N+1}/N$
in the thermodynamic limit $N\to\infty$, in accordance with Remark \ref{rem:n:n}.
This is also true for an interaction where $j(d_0)>0$ for some $d_0$ with 
${\rm gcd}(d_0,N)=1$ and $j(d)=0$ otherwise. \hfill $\Diamond$
\end{example}
\begin{remark}[Genericity of stable degeneracy] \label{rem:generic}
As the maps $j\mapsto H(\sigma,j)$ are linear, $j$--degeneracy
$D(\sigma,j)$  (defined in \eqref{j:degeneracy}) equals $D_{\rm stab}(\sigma)$ for
all $j$ in the complement of a finite union of subspaces in $J$
that are of codimension one. 

So joint stability of all degeneracies 
is open and dense and of full Lebesgue measure in the parameter space $J$. 
\hfill $\Diamond$
\end{remark}
\begin{remark}[Exterior field] 
An additional translation invariant coupling to the exterior magnetic
field $h\in\IR$ in the energy function \eqref{energy} leads to
\[\widetilde{H}\equiv  \widetilde{H}_F:G_F\times J_F\times \IR_h
\to \IR\qmbox{,} 
\widetilde{H}_F(\sigma,j,h)=\langle (j,h), \tilde{A}(\sigma)\rangle,\]
with the modified correlation map
\beq
\tilde{A}_F:G_F\to \IZ^F\times\IZ\qmbox{,} \tilde{A}_F(\sigma)
=\Big(A_F(\sigma),\sum_{f\in F}\sigma_f\Big). 
\Leq{modif}
Then $\tilde{A}$ and thus $\widetilde{H}$ is still invariant under 
translations and reflections.

The redefined stable degeneracy (compare with \eqref{Dstab})
\[\tilde{D}_{\rm stab}(\sigma)
:=\big| \tilde{A}_F^{-1}\big(\tilde{A}_F(\sigma)\big) \big| \] 
equals ${D}_{\rm stab}(\sigma)$ iff
$|\{f\in F\mid \sigma_f=1\}|=|F|/2$, and
equals $\eh D_{\rm stab}(\sigma)$ otherwise.
The reason is the equality
\[\Big(\sum_{f\in F}\tau_f\Big)^2=\sum_{k\in F} A(\tau)_k
=\sum_{k\in F} A(\sigma)_k=\Big(\sum_{f\in F}\sigma_f\Big)^2\]
in the case $A(\tau)=A(\sigma)$. So the last term in \eqref{modif}
is determined by the first term up to sign, which equals 0 iff
$|\{f\in F\mid \sigma_f=1\}|=|F|/2$.
\hfill $\Diamond$
\end{remark}
%
\section{Configurations With Large Stable Degeneracy} \label{sect:3}
%
In this section we start our search for spin configurations $\sigma$ 
where $D_{\rm stab}(\sigma)>D_{\rm sym}(\sigma)$.
Generally speaking, our examples are based on different
kinds of 'hidden symmetries' of~$\sigma$. 

\begin{example}[Product configurations]  
A rather trivial case concerns the degeneracy of a product configuration
\eqref{product:conf}, whose factors $\sigma^{(j)}\in G_{F_j}$ have the 
maximal symmetry-induced degeneracy, that is 
$D_{\rm sym}(\sigma^{(j)})=4|F_j|$. 

Then $D_{\rm sym}(\sigma)=4|F|$,
but by Prop.\ \ref{prop:prod} \ 
$D_{\rm stab}(\sigma)\ge 2^{1-d}\prod_{j=1}^d D_{\rm stab}(\sigma^{(j)})
\linebreak 
\ge 2^{1-d}\prod_{j=1}^dD_{\rm sym}(\sigma^{(j)})$
$=2^{d+1}|F|$. So in this case 
$D_{\rm stab}(\sigma)\ge 2^{d-1}D_{\rm sym}(\sigma)$.
In other words the quotient 
$D_{\rm stab}(\sigma)/D_{\rm sym}(\sigma)\ge 1$ is unbounded in general.
\hfill $\Diamond$
\end{example}

\begin{figure}[t]
\begin{center}
\includegraphics[draft,height=.5\textwidth,natwidth=360,natheight=375]
{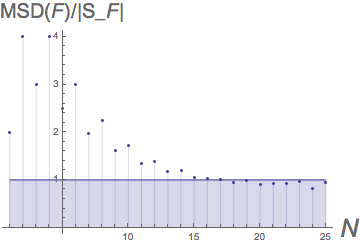}
\caption{The quotient ${\rm MSD}(F)/ |{\mathcal S}_F|$ for $F=\IZ/N\IZ$, 
see Conjecture \eqref{conjecture}}
\label{fig:conj}
\end{center}
\end{figure}

One strategy to find more interesting configurations $\sigma$ with large
stable degeneracies
is based on the equivariance property \eqref{equivar}. 
We are seeking spin configurations $\sigma\in G_F$ and automorphisms 
$g\in {\rm Aut}(F)$ so that $A_F(\sigma)$ is invariant under 
$\Psi^{({\rm ev})}_g$, but
$\Psi^{(G)}_g(\sigma)$ is not in the $\Phi$--orbit of $\sigma$.
The correlation $A_F(\sigma):F\to\IZ$ should not have a
large image in $\IZ$ to allow for such $g\in {\rm Aut}(F)$.
Such configurations $\sigma$ are of some interest, independent of
whether they lead to a large stable degeneracy.
The only case with $|A_F(\sigma)|=1$ is $\sigma=\pm \idty_F$. 
In the examples below, $|A_F(\sigma)|\le3$ (Prop.\ \ref{prop:Legendre}),
respectively $|A_F(\sigma)|=2$ (Prop.\ \ref{prop:PDS}).
\begin{remark}[Empirical data] \label{rem:empirical}
We performed a computer search for 
degeneracies of the spin configurations $\sigma\in G_F$
with $F=\IZ/N\IZ$ and integers $N\le 15$.
In analyzing the data, a large variety of phenomena was found. Some examples:
\begin{itemize}
\item 
The first case where the inequality $D_{\rm sym}(\sigma)\le|{\mathcal S}|=4N$ 
(see \eqref{D:sym}) is saturated, occurs for $N=7$ and $\sigma=(1,1,-1,1,-1,-1,1)$.
It has the unusual property
$A_F(\sigma)_f=-1$ $(f\in F\setminus\{0\})$, 
but $D_{\rm stab}(\sigma)=D_{\rm sym}(\sigma)=4N$.
This will be explained in number-theoretic terms  (Example \ref{ex:Leg7}).
\item 
The first $\sigma$ whose stable degeneracy exceeds its
symmetry-induced degeneracy has length $N=12$, and 
$D_{\rm stab}(\sigma)=2D_{\rm sym}(\sigma)=8N$.
That stable degeneracy follows from the action of ${\rm Aut}(F)$.
\item 
The stable degeneracy $D_{\rm stab}(\sigma)=2|{\mathcal S}|=8N$ is also found
for certain configurations $\sigma$ with $N=13$ to $15$.
We will explain the case $N=13$ using the projective plane ${\rm PG}(2,3)$ (Example \ref{ex:N13}).
\item 
$N=14$ is interesting in that  
\[\sigma:=(-1, -1, 1, 1, 1, 1, 1, -1, 1, 1, -1, 1, -1, 1)\in F\] 
has stable 
degeneracy $D_{\rm stab}(\sigma)=2D_{\rm sym}(\sigma)=8N$, and
\[\tau:=(-1, -1, 1, -1, 1, 1, 1, 1, -1, 1, 1, 1, -1, 1)\] 
belongs to the same class,
but $\tau$ is not in the orbit of the action \eqref{eq:joint}.
\item 
For $N=16$, $\sigma:=(-1,-1, 1, -1, 1, 1, 1, 1, -1, 1, -1, 1, -1, -1, 1, 1)$
has stable degeneracy $D_{\rm stab}(\sigma)=3|{\mathcal S}|=12N$.
${\rm Aut}(\IZ/16\IZ)$ maps $A(\sigma)$ to a four-element set of
correlations, clearly with the same  stable degeneracy. 
\end{itemize}

Conjecture \eqref{conjecture} predicts that the mean stable degeneracy
is asymptotic to the average symmetry induced degeneracy.
This is compatible with the data of Figure~\ref{fig:conj}.
\hfill $\Diamond$
\end{remark}

We begin with a number-theoretic construction of certain configurations $\sigma\in G_F$. 
This is interesting as the correlation of these $\sigma$ takes only two or three values, and
in fact $\Psi^{({\rm ev})}_g \big(A_F(\sigma)\big) = A_F(\sigma)$ for all $g\in{\rm Aut}(F)$.  
But as $\Psi^{(G)}_g(\sigma)\in\Phi({\mathcal S},\sigma)$, the construction does not
lead to large stable degeneracy (see Rem.~\ref{rem:deg:Legendre}).

\begin{prop}[Correlation for Legendre symbols]  \label{prop:Legendre}
\hspace*{-2mm} For primes $N\in\IP\setminus\!\{2\}$ and $F=\IZ/N\IZ$ the group elements 
$\sigma^\pm\in G_F$, given by the values
$\sigma^\pm_k:=\left(\frac{k}{N}\right)$ of the Legendre symbol
for $k=1,\ldots,N-1$ and $\sigma^\pm_N:=\pm1$, have correlations

\begin{enumerate}
\item[$\bullet$] 
$A_F(\sigma^\pm)_f=-1$ for all $f\in F\setminus\{0\}$, if $N\equiv 3\!\! \mod4$. 
\item[$\bullet$] 
$A_F(\sigma^\pm)_f=\left(-1+2\sigma^\pm_N\left(\frac{f}{N}\right)\right)$ for all 
$f\in F\setminus\{0\}$, if $N\equiv 1\!\! \mod4$.
\end{enumerate}
\end{prop}
\noindent
\textbf{Proof:}
$\bullet$ 
For $N\equiv 3\!\! \mod4$ and $f\in F\setminus\{0\}$ we express the correlation
entirely in terms of Legendre symbols. This is possible, since 
$\left(\frac{0}{N}\right)=0$, and the two terms in $A_F(\sigma^\pm)_f$
involving $\sigma^\pm_N$ cancel, as 
$\left(\frac{-f}{N}\right)=\left(\frac{-1}{N}\right)\left(\frac{f}{N}\right)$ and 
$\left(\frac{-1}{N}\right)=(-1)^{(N-1)/2}=-1$. So
\[A_F(\sigma^\pm)_f = \sum_{\ell\,\in\, \IZ/N\IZ} \sigma^\pm_\ell\sigma^\pm_{f+\ell}
= \sum_{\ell\,\in\, \IZ/N\IZ}\ \left(\frac{\ell}{N}\right)\left(\frac{f+\ell}{N}\right)
= \sum_{\ell\,\in\, \IZ/N\IZ}\ \left(\frac{(f+\ell)\ell}{N}\right).\]
We used here that $a\mapsto\left(\frac{a}{b}\right)$ is completely multiplicative.
Now the equation $x^2=(f+\ell)\ell$ has $N-1$ solutions $(x,\ell)\in (\IZ/N\IZ)^2$.
This follows by the substitution $a:=x+\ell+f/2$, $b:=x-\ell-f/2$, which implies
$ab=x^2-(f+\ell)\ell-f^2/4$. But since then $\left(\frac{(f+\ell)\ell}{N}\right)=0$
if $x=0$ and $\left(\frac{(f+\ell)\ell}{N}\right)=1$ otherwise,
this number of solutions also equals 
$ \sum_{\ell\,\in\, \IZ/N\IZ}  \left(1+\left(\frac{(f+\ell)\ell}{N}\right)\right)
=N+\sum_{\ell\,\in\, \IZ/N\IZ}\ \left(\frac{(f+\ell)\ell}{N}\right)$.\\
$\bullet$ 
For $N\equiv 1\!\! \mod4$ we obtain the additional term
$\sigma^\pm_N (\sigma^\pm_f+\sigma^\pm_{-f})
=2\sigma^\pm_N\left(\frac{f}{N}\right)$ 
in $A_F(\sigma^\pm)_f$.
\hfill $\Box$
\begin{remark}[Degeneracy for Legendre symbols] \label{rem:deg:Legendre}
The automorphism group
\[{\rm Aut}(\IZ/N\IZ) = \{a\in \IZ/N\IZ\mid {\rm gcd}(a,N)=1\}\] 
acts by multiplication on $\IZ/N\IZ$. We consider
$\sigma^\pm$  from Proposition \ref{prop:Legendre}.
\begin{itemize}
\item 
In the case $N\equiv 3\!\! \mod4$ the $\Phi$--orbits
through $\sigma^+$ and $\sigma^-$ coincide, as 
$\sigma^\mp=\Phi_{(-1,0,-1)}(\sigma^\pm)$.
That orbit is left invariant by $\Psi^{(G)}_k$
(since for $k\in {\rm Aut}(\IZ/N\IZ)$, $\Psi^{(G)}_k(\sigma^\pm)=\sigma^\pm$ 
for residues $k$ and $\Psi^{(G)}_k(\sigma^\pm)=\Phi_{(1,0,-1)}(\sigma^\pm)$ 
for non-residues $k$).
\item 
For $N\equiv 1\! \! \mod4$
the $\Phi$--orbits through $\sigma^+$ and $\sigma^-$
are different, since 
$A_F(\sigma^+)\neq A_F(\sigma^-)$, see Proposition \ref{prop:Legendre}.
The automorphisms $\Psi^{(G)}_k$ leave these orbits invariant for
residues $k$ and interchanges them for non-residues $k$
(as $\Psi^{(G)}_k(\sigma^\pm)=\sigma^\pm$, respectively
$\Psi^{(G)}_k(\sigma^\pm)=\Phi_{(-1,0,1)}(\sigma^\mp)$). 
\end{itemize}
So in both cases
we cannot conclude that $D_{\rm stab}(\sigma^\pm)$ is strictly larger
than $D_{\rm sym}(\sigma^\pm)$.
\hfill $\Diamond$
\end{remark}
\begin{example}[Legendre symbols]  \label{ex:Leg7}
For $F=\IZ/7\IZ\cong\{1,\ldots,7\}$, the configuration
$\sigma=(1,1,-1,1,-1,-1,1)\in G_F$ of Remark \ref{rem:empirical}
equals $\sigma^+$.
But
$\sigma$ can also be understood in terms of the Fano plane 
${\rm PG}(2,2)$, see Section \ref{sect:4}.
\hfill $\Diamond$
\end{example}
%
\section{Stable Degeneracy and Singer Sets} \label{sect:4}
%
We continue our search for spin configurations with large stable degeneracy.

For the prime power $q:=p^n$ ($p\in\IP$ and $n\in\IN$) the Desarguesian
projective plane ${\rm PG}(2,q)$ consists of $N := q^2+q+1$ elements 
({\em points}).
These are the one-dimensional subspaces
of the vector space $\IF_q^3$ over the Galois field $\IF_q$.
The {\em lines} of ${\rm PG}(2,q)$ are the two-dimensional subspaces of 
$\IF_q^3$
and are considered as subsets of ${\rm PG}(2,q)$. So there are $N$ lines, too.

As will be shown below,
${\rm PG}(2,q)$ allows the construction of configurations 
$\sigma\in G_F$ for $F=\IZ/N\IZ$ 
with constant $A_F(\sigma)_f = 1$ ($f\in F\setminus\{0\}$) 
and large stable degeneracy $D_{\rm stab}(\sigma)$. 

Automorphisms (called {\em collineations}) of ${\rm PG}(2,q)$
are bijections, mapping lines to lines.
According to the fundamental theorem of projective geometry, they
are induced by bijective {\em semilinear} maps $\phi:\IF_q^3\to \IF_q^3$, 
that is, for some automorphism $\tau\equiv \tau_\phi:\IF_q\to \IF_q$
\[ \phi(v+w)=\phi(v)+\phi(w)\qmbox{and} \phi(\lambda v)=\tau(\lambda)\phi(v)
\qquad \big(v,w\in\IF_q^3,\, \lambda\in\IF_q\big).\]
As $\phi$ maps $k$--dimensional subspaces to $k$--dimensional subspaces, 
it descends to an automorphism $\tilde \phi\in {\rm Aut}({\rm PG}(2,q))$.

As a vector space over $\IF_q$, $\IF_q^3\cong\IF_{q^3}$.
The {\em Singer subgroup} $\Sigma$ of ${\rm Aut}({\rm PG}(2,q))$
consists of the automorphisms
induced by multiplication with the non-zero elements of $\IF_{q^3}$.
It is thus cyclic and of order $N$ 
(see Hughes and Piper \cite{HP} and the original article \cite{Si} by Singer).
Concretely let the irreducible primitive cubic $X^3-c_2X^2-c_1X-c_0$, 
$c_i\in\IF_q$ define multiplication in $\IF_{q^3}$.
For a root $\lambda\in\IF_{q^3}$ of that cubic
\[\lambda^3=c_2\lambda^2+c_1\lambda+c_0,\] 
so that multiplication with $\lambda$ corresponds in the ordered basis 
$(\lambda^2,\lambda,1)$ of $\IF_q^3$ to multiplication with
the matrix 
\[M :=
\left(\begin{smallmatrix}c_2&1&0\\c_1&0&1\\c_0&0&0\end{smallmatrix}\right)
\in{\rm GL}(3,\IF_q).\]
$M$ gives rise to a projective collineation
 $\widetilde M:{\rm PG}(2,q)\to {\rm PG}(2,q)$
of period $N$. 
We use additive notation, identifying $\Sigma$ with $\IZ/N\IZ$.
Starting with a point $A_0\in{\rm PG}(2,q)$ we denote 
the points of its orbit ${\rm PG}(2,q)$ by 
\[A_k:=\widetilde M^k(A_0)\qquad(k=0,\ldots, N-1).\]
For a point $A$ and a projective line $L$ in ${\rm PG}(2,q)$ their 
{\em difference set} is defined as 
\[ {\mathcal D} \equiv{\mathcal D}(A,L) := \{g\in \Sigma\mid g(A)\in L\}.\]
Its cardinality thus equals the one of the projective  line: $|{\mathcal D}| = q+1$
(see also Golay \cite{Go}).

This set of group elements is {\em perfect}:
for every $h\in \IZ/N\IZ\setminus\{0\}$ there are unique $d_1,d_2\in {\mathcal D}$
with $h=d_2-d_1$ (see Lemma 13.12 of \cite{HP}).
If we attribute to a subset ${\mathcal D}\subseteq \IZ/N\IZ$ the spin
configuration
\[(-1)^{\idty_{\mathcal D}} \in G_F,\]
then for a perfect difference set ${\mathcal D}$ it is mapped by \eqref{A:N}
to 
\beq
A\big( (-1)^{\idty_{\mathcal D}} \big)_0=N\qmbox{,}
A\big( (-1)^{\idty_{\mathcal D}} \big)_f = q^2-3q+1 \quad
(f\in F\setminus\{0\}).
\Leq{Apds}

Assuming $L$ to be the projective line through $A_0$ and $A_1$,
${\mathcal D}(A_0,L)$ contains $0$ and $1\in \IZ/N\IZ$.
There is a unique shift that lets a perfect difference set contain $0$ and $1$, 
and the 
corresponding set is called {\em reduced} in  \cite{Si}.
This normalization allows us to discern difference sets not just being 
mutual translates in $\IZ/N\IZ$.
\begin{prop}  \label{prop:PDS}
For the perfect difference sets ${\mathcal D}(A,L)$ the stable degeneracy 
is bounded by 
$D_{\rm stab} \big((-1)^{\idty_{\mathcal D}} \big) \ge 2N\varphi(N)/(3n)$.
\end{prop}
\noindent
\textbf{Proof:}
A conjecture in \cite{Si} claims that there
are exactly $\varphi(N)/(3n)$ reduced perfect difference sets.
This conjecture has been proven 25 years later by Halberstam and Laxton in
\cite{HL}.
For perfect difference sets ${\mathcal D}$, 
$D_{\rm sym} \big((-1)^{\idty_{\mathcal D}} \big)  = 4N$, the maximal 
value possible for an element of $G_F$. This follows, since
\begin{itemize}
\item 
the period of $(-1)^{\idty_{\mathcal D}}$ equals $N$, because otherwise 
${\mathcal D}$ could not be perfect;
\item 
$(-1)^{\idty_{\mathcal D}}$ is not palindromic, for the same reason,
\item 
The number of entries $-1$ in $(-1)^{\idty_{\mathcal D}}$ equals
$q+1<N/2$, since $N=q^2+q+1$ and $q=p^n\ge2$. So $-(-1)^{\idty_{\mathcal D}}$
is not representable in the form $\Phi_{(1,t,s)}\big((-1)^{\idty_{\mathcal D}}\big)$.
\end{itemize}
The number of is only half of the product 
of $\big|D_{\rm sym} \big((-1)^{\idty_{\mathcal D}} \big) \big|$ 
and $\varphi(N)/(3n)$, since the reflections $\psi_{-1}=\Phi_{(1,0,-1)}$ 
are counted twice in that product.
\hfill $\Box$\\[2mm]
Stable degeneracy of $\sigma\in G_F$ can be nearly as large as 
the square of the group order of $F$, that is, 
as the square of symmetry-induced degeneracy:
\begin{corollary}
For all $q\in \IP$, $N:=q^2+q+1$ there is a configuration 
$\sigma \in G_{\IZ/N\IZ}$ with
\beq
D_{\rm stab}(\sigma)\ge \frac{N^2}{3\log(\log N)}.
\Leq{d:nn}
\end{corollary}
\noindent
\textbf{Proof:}
We use the lower bound $2N\varphi(N)/3$, with $N:=q^2+q+1$ and
$q\in\IP$. 
However, we can use that 
(see Bach and Shallit \cite[Theorem 8.8.7]{BS})  for $m\ge 3$
$\varphi(m)\ge m/\big(e^\gamma \log(\log m)+3/\log(\log m) \big)$.
The right hand side is smaller than $m/\big(2 \log(\log m))$ for 
all large $m$, and we tested \eqref{d:nn} for the first $35\,000\,000$ 
cases $m=N(q)$ with $q\in\IP$, which are not covered by that inequality.
\hfill $\Box$
\begin{remark}[Stable degeneracy quadratic in the group order]
\ \ 
It is unknown (see Baier and Zhao \cite[Section 1]{BZ})
whe\-ther for any quadratic polynomial $P\in \IZ[q]$
there are infinitely many primes in $P(\IN)$. So it is unclear, 
whether for some $c>0$
there are infinitely many prime powers $q$ with $\varphi(N)\ge c N$
for $N=q^2+q+1$.

Thus we do not know how to prove that for some $c\in(0,2/3)$ we have
$D_{\rm stab}(\sigma)\ge c N^2$ for infinitely many $N$ 
and some $\sigma \in G_{\IZ/N\IZ}$.
\hfill $\Diamond$
\end{remark}
\begin{example}[Large stable degeneracy by Singer sets, for $F=\IZ/13\IZ$] 
\ \label{ex:N13} 

\noindent
We consider the group $F=\IZ/N\IZ\cong\{0,1,\ldots,12\}$ for $N=13$.  
\[\sigma=(1, 1, -1, 1, -1, -1, -1, -1, -1, 1, -1, -1, -1)\in G_F\]
has the degeneracies $D_{\rm sym}(\sigma)=|{\mathcal S}_F|=4N$ and
$D_{\rm stab}(\sigma)=8N$. The explanation is the following:
For $q=3$, $N=q^2+q+1$, and  $\sigma = (-1)^{\idty_{\mathcal D}}$ 
corresponds to a Singer set ${\mathcal D}\subset F$ 
of the projective plane ${\rm PG}(2,q)$. 
By \eqref{Apds} its correlation equals 
\[ A\big( \sigma \big)_0=13 \qmbox{, and} 
A\big( \sigma \big)_f = 1 \quad \big(f\in (\IZ/13\IZ)\!\setminus\!\{0\}\big).\]
So the correlation is constant on the $\Psi$-orbit of $\sigma$. 
$\Psi(3,\sigma)$ and $\Psi(9,\sigma)$ are translates of $\sigma$, 
whereas $\Psi(4,\sigma)$, $\Psi(10,\sigma)$ and $\Psi(12,\sigma)$ 
additionally reflect it. On the other hand,
$\Psi(2,\sigma)=(1,1,-1,-1,1,-1,1,-1,-1,-1,-1,-1,-1)$ 
 is not in the $\Phi$-orbit of $\sigma$, leading to 
 $D_{\rm stab}(\sigma) = 2 D_{\rm sym}(\sigma)$. 
Here the bound of Proposition \ref{prop:PDS} is sharp, as 
$2N\varphi(N)/(3n)= 8N=104$.

Incidentally, this $\sigma\in G_F$ is also defined by $\sigma_k:=1$ if $k=x^4$ 
for some $x\in \IZ/N\IZ$ and  $\sigma_k:=-1$ otherwise.
\hfill $\Diamond$
\end{example}

\begin{remark}[Block designs] 
The notion of a Singer group is generalized to be an automorphism group
acting regularly on the blocks of a symmetric block design. Then
every such symmetric block design has a representation by 
a difference set, see Theorem XI 5.2 of Jacobs and Jungnickel \cite{JJ}.
\hfill $\Diamond$
\end{remark}
Using this and similar ideas one can probably construct many spin configurations
with large stable degeneracy.
%
\section{Substitutions} \label{sect:5}
%
We now apply a technique, borrowed from trace identities and going back to
Horowitz \cite{Ho},
to find more configurations of large stable degeneracy.

Over a finite, nonempty set ${\mathcal A}$ (called {\em alphabet} with {\em
letters} $a\in  {\mathcal A}$),
\[{\mathcal A}^*:=\epsilon\,\cup\,\textstyle \bigcup_{n\in \IN}{\mathcal A}^n\]
is the set
of finite {\em words}, with $w\in {\mathcal A}^n$ a word of {\em length} 
$|w|:=n$, and $\epsilon$ the {\em empty word} of length $|\epsilon|:=0$. 
With {\em concatenation}
$(v_1\ldots v_m,w_1\ldots w_n)\mapsto v_1\ldots v_mw_1\ldots w_n$,
${\mathcal A}^*$ becomes a monoid with identity $\epsilon$.
For $a\in {\mathcal A}$ and $w\in {\mathcal A}^*$, $|w|_a\in \IN_0$
denotes the number of occurrences of the letter $a$ in $w$.
The {\em inverse} of $v:=v_1\ldots v_m\in {\mathcal A}^*$ is 
$v^{-1}:=v_m\ldots v_1$, and in the case ${\mathcal A}=\{-1,1\}$
their correlations coincide:
\beq
A_{\IZ/m\IZ}(v^{-1})=A_{\IZ/m\IZ}(v).
\Leq{AF:inv}
Given words $U,V\in {\mathcal A}^*$ over the alphabet ${\mathcal A}:=\{-1,1\}$
and a word $W\equiv W(U,V)\in \{U,V\}^*$, by concatenation
one considers $W$ as a word $\widetilde W \in{\mathcal A}^*$, and thus
as a configuration $\sigma\in G_F$ for some group $F=\IZ/N\IZ$. Here 
$|\widetilde W|_{\pm1}=|W|_U \,|U|_{\pm1}+|W|_V \, |V|_{\pm1}$
and therefore $N=|W|_U \,|U| +|W|_V \, |V|$.

Then $X:= W(U,V)^{-1}$ gives rise to
$\widetilde X \in{\mathcal A}^*$. This does {\em not} in general
coincide with $(\widetilde W)^{-1}$.
\begin{prop}[Correlation map and substitutions]  \label{prop:subs}
$A_F(\widetilde X)=A_F(\widetilde W).$
\end{prop}
\noindent
\textbf{Proof:}
For the word $W=w_1\ldots w_\ell$ over the alphabet $\{U,V\}$
the reversed word equals $X=w_\ell\ldots w_1$. We use
cyclic indexing, that is $1,\ldots,\ell\in F':=\IZ/\ell\IZ$. 
\begin{enumerate}[1.]
\item 
Thus in $W$ and $X$ for any $a,b\in F'$ the letters $w_a$, $w_b$ 
have the same cyclic distance.
Moreover, they are separated by the same letters $w_{a+1},\ldots,w_{b-1}$
respectively $w_{b+1},\ldots,w_\ell,w_1,\ldots,w_{a-1}$.

So if $w_a=w_b$, then the contribution of the pair in
$A_F(\widetilde W)$ coincides with the one in $A_F(\widetilde X)$.
\item 
Consider now pairs $w_a\neq w_b$, say, $w_a=U$ and $w_b=V$. Then for all
$k\in \IZ/\ell\IZ$ there is a bijection $\rho_k:P_k\to Q_k$ between the two sets
\[P_k := \{c\in F'\mid w_c=U, w_{c+k}=V \}\ ,\
Q_k:=\{d\in F'\mid w_d=V, w_{d+k}=U \},\]
with frequencies of separating letters $U$
\[\big(w_{\rho_k(c)+1}\ldots w_{\rho_k(c)+k-1}\big)_U=
\big(w_{c+1}\ldots w_{c+k-1}\big)_U\]
(and the same for $V$).
This follows by induction in the length $\ell'$ of the subword $w_1\ldots w_{\ell'}$,
$\ell'=1,\ldots,\ell$.
So the contribution of the pairs $(U,V)$ in
$A_F(\widetilde W)$ coincides with the one in $A_F(\widetilde X)$,
and similarly for the pairs $(V,U)$.
\hfill $\Box$
\end{enumerate}
\begin{example}[Substitution]  \label{ex:subs}
$W:=UVUUVVV$, with $U:=(1,1,-1)$ and $V:=(-1,1,-1)$. So for $F:=\IZ/21\IZ$
the configuration equals $\widetilde W=\sigma \in G_F$, with
\[\sigma:=(1,1,-1, -1,1,-1, 1,1,-1, 1,1,-1, -1,1,-1, -1,1,-1, -1,1,-1).\]
The correlation of $\sigma$ equals $A(\sigma)_0=21$,
and for $f\in F\setminus\{0\}$: \\
$A(\sigma)_f=13$ if $f= 0\;(\!\!\!\mod3)$ and $A(\sigma)_f=-7$
otherwise.
Thus, unlike $\sigma$, $A(\sigma)$ is a fixed point of the ${\rm Aut}(F)$ 
action.\\[2mm]
The inverse of $W$ equals 
$X=VVVUUVU$. So $\widetilde X=\tau\in G_F$, with
\[\tau:=(-1,1,-1,-1,1,-1,-1,1,-1,1,1,-1,1,1,-1,-1,1,-1,1,1,-1 ).\]
$\tau$ is a translate of $\Psi^{(G)}_{10}(\sigma)$.
It is not an element of the orbit $\Phi({\mathcal S},\sigma)$:
\begin{itemize}
\item 
As $|\sigma|_1=|\tau|_1=10\neq11=|\sigma|_{-1}$ $=|\tau|_{-1}$,
if $\Phi\big((s,t,r),\sigma\big)$ there can be no spin flip ($s=1$).
\item 
In addition a pure translation ($r=1$) is impossible, since the 
subsequence $(-1,1,1,-1,1,1,-1,-1,1,-1,-1)$ of $\sigma$ is not a subsequence of
the cyclic word $\tau$. 
\item 
It cannot be a reflection ($r=-1$) either, since $\tau$ does not contain
the subsequence $(-1,1,-1,-1,1,-1,-1,1,-1,-1)$ of $\Phi\big((1,0,-1),\sigma\big)$.
\end{itemize}
So $D_{\rm stab}(\sigma) > D_{\rm sym}(\sigma)$. 
\hfill $\Diamond$
\end{example}
The method of Proposition \ref{prop:subs} can be iterated.
%
\section{Block Sizes and Stable Degeneracy}  \label{sect:6}
%
The discrete Laplacian of the correlation reveals some
information about the spin configurations sharing that correlation,
in particular the structure of blocks of alike spins.
This may be useful regarding Conjecture \eqref{conjecture}.
\begin{lemma}[Laplacian for correlations]  \label{lem:laplace}
For the residue class group $F:=\IZ/N\IZ$, $N\in \IN$ the Laplacian 
$\Delta: \IR^F\to \IR^F$, 
$(\Delta h)_f={\textstyle \frac{1}{4}} (h_{f-1}-2h_{f}+h_{f+1})$
is injective as a map $A_F(G_F)\to \IZ^F\cap \IR^F_{\rm ev}$.
\end{lemma}
\textbf{Proof:}
The normalization $1/4$ is chosen so that by the first statement in
Prop.\ \ref{prop:image} the restriction of $\Delta$
to the image of the correlation map has its image in $\IZ^F$.
Moreover, $A_F(G_F)\subseteq \IR^F_{\rm ev}$ and
$\Delta$ restricts to  $\IR^F_{\rm ev}$.\\
Injectivity follows, since ${\rm ker}(\Delta)=\IR \idty_F$, and 
$A_F(\sigma)_0=N$ for all $\sigma\in G_F$.
\hfill $\Box$\\[2mm]
So in principle one can calculate the stable degeneracy of a configuration
$\sigma\in G_F$ by considering $\Delta A_F(\sigma)$.

The only configurations with constant correlation are $\pm\idty_F$ (since then
$A_F(\sigma)_f$ $= A_F(\sigma)_0 = N$).
For a residue class group $F=\IZ/N\IZ$ the other configurations
are thus translates of $\sigma\in G_F$ which are of the form
\beq 
\sigma = (1)^{m_1}(-1)^{m_2}\ldots(1)^{m_{2k-1}}(-1)^{m_{2k}}
\Leq{sigma:2m}
with $k\in\IN$ and $m_\ell\in \IN$ $(\ell=1,\ldots,2k)$, and
the strings $(1)^m$ (respectively $(-1)^m$) 
of $m$ ones (respectively $-1$).
So $\sum_{\ell=1}^{2k} m_\ell=N$.
Note that stable and symmetry-induced degeneracy are invariant under 
translation of a configuration. Thus it is natural to consider 
the indices $j$ of $m_j$ as elements of $\IZ/(2k\IZ)$.
\begin{lemma}  \label{lem:Delta:A}
For $F:=\IZ/N\IZ$ and a configuration $\sigma\in G_F$ 
of the form \eqref{sigma:2m},
\beq
\Delta\; A_F(\sigma) = \sum_{\ell=1}^{2k} \sum_{n=1}^k 
\big(
\delta_{m_\ell+\ldots+m_{\ell+2n-2}}-
\delta_{m_\ell+\ldots+m_{\ell+2n-1}}
\big).
\Leq{eq:laplace:block}
\end{lemma}
\textbf{Proof:}
For all $f\in F$, \ $\Delta\, A_F(\sigma)_f 
= {\textstyle \frac{1}{4}}  
(A_F(\sigma)_{f-1} - 2A_F(\sigma)_{f} +A_F(\sigma)_{f+1}) = $
\[ {\textstyle \frac{1}{4}}\!\! \sum_{g\in\IZ/N\IZ} \sigma_g
(\sigma_{g+f-1}-2\sigma_{g+f}+\sigma_{g+f+1})
= \!\!  \sum_{g\in\IZ/N\IZ} \!\!{\textstyle \frac{1}{4}}
(\sigma_{g}-\sigma_{g-1})(\sigma_{g+f-1}-\sigma_{g+f}).\]
The terms $\frac{1}{4}(\sigma_{g}-\sigma_{g-1})(\sigma_{g+f-1}-\sigma_{g+f})$
in the last sum are non-zero iff $\sigma_{g}\neq\sigma_{g-1}$
and $\sigma_{g+f-1}\neq\sigma_{g+f}$, that is, iff the indices are
of the form $g=\sum_{\ell=1}^r m_\ell$ and 
$g+f=\sum_{\ell=1}^s m_\ell$ for some $r,s$. The modulus of these terms
equals one, and their sign is negative iff $s-r$ is even. This explains the
$\delta$--terms in \eqref{eq:laplace:block}.
\hfill $\Box$
\begin{remarks}[Block sizes] \label{rem:reconstuct}
\begin{enumerate}
\item 
Note that Lemma \ref{lem:Delta:A} implies that
we can read off the number $2k$ of blocks in the configuration
\eqref{sigma:2m} from the Laplacian of its correlation: 
$\big(\Delta\, A_F(\sigma)\big)_0=-2k$.
\item 
In the substitution example~\ref{ex:subs}
the multisets of block sizes for $\sigma$ and $\tau$ both equal $1^72^7$
($1^42^3$ for the blocks of ones, and $1^32^4$ for the blocks of minus ones).

However, the Singer set in Example \ref{ex:N13} shows that the correlation does not 
in general determine the multiset of block sizes $m_\ell$:
$\sigma$ has block sizes $(m_1, \ldots,m_6)=(2,1,1,5,1,3)$
and multiset $1^32^13^15^1$, which differs from the multiset $1^32^26^1$
for the data $(2,2,1,1,1,6)$ of $\tau:=\Psi(2,\sigma)$ with 
$A_F(\sigma)=A_F(\tau)$.
\item 
Nevertheless, the multiplicity of the {\em minimal} block size
$\min(m_1, \ldots,m_{2k})$ can be deduced from the correlation
and equals
$\big(\Delta\, A_F(\sigma)\big)_\ell$, with $\ell>0$ the smallest index for
which $\big(\Delta\, A_F(\sigma)\big)_\ell\neq 0$.
\hfill $\Diamond$
\end{enumerate}
\end{remarks}
From Remark \ref{rem:reconstuct}.2 we see that non-trivial stable degeneracy
is possible for $2k=6$--block configurations. We show now that this is the minimal
number. The method is to reconstruct
for $k=1$ and $k=2$ from $\Delta\, A_F(\sigma)$ 
the list $(m_1,\ldots, m_{2k})$ of block sizes, up to cyclic permutations and reflection.
It will turn out that already for $k=2$ the combinatorics is intricate. To simplify the
proof, we use the following observation.

\begin{lemma} [$\ell^1$--norm] \label{lem:l1:norm}
All $2k$--block configurations $\sigma\in G_F$ fulfill the inequalities
\beq
4k \le \|\Delta\, A_F(\sigma)\|_1 \le 4k^2,
\Leq{ineq:l1}
and $\|\Delta\, A_F(\sigma)\|_1$ is a multiple of four.
\end{lemma}
\textbf{Proof:}
\eqref{ineq:l1} is true for $k=0$, that is, $\sigma=\pm \idty$,  since 
$\Delta\, A_F(\pm \idty_F)=0$.
\newline
For $\sigma\in G_F\setminus\{-\idty_F,\idty_F\}$ we obtain the right inequality in \eqref{ineq:l1}
by counting the number $4k^2$ of terms in \eqref{eq:laplace:block}.
The left inequality follows from the observation that $(\Delta\, A_F(\sigma))_0=-2k$
and that $\sum_{f\in F}(\Delta\, A_F(\sigma))_f=0$.
$\|\Delta\, A_F(\sigma)\|_1$  is a multiple of four, since $\Delta\, A_F(\sigma)$ is an even function (Lemma \ref{lem:laplace}),
and since even and odd contributions to \eqref{eq:laplace:block} cancel in pairs.
\hfill $\Box$\\[2mm]
On the r.h.s.\ of \eqref{ineq:l1} equality occurs if and only if there is no cancellation between
$\delta$--functions with different signs (that is, there are no 
$1\le \ell_1,\ell_2\le 2k$ and $1\le n_1,n_2\le k$ with 
$m_{\ell_1}+\ldots+m_{\ell_1+2n_1-1} = m_{\ell_2}+\ldots+m_{\ell_2+2n_2}$).
Then given $\Delta\, A_F(\sigma)$, we know the multisets of lengths 
$m_{\ell_1}+\ldots+m_{\ell_1+2n_1-1}$ respectively 
$m_{\ell_2}+\ldots+m_{\ell_2+2n_2}$ of unions of adjacent blocks.
In addition we then know whether for a given element of the multiset 
the number of blocks is even or odd. 
\begin{prop}[Trivial stable degeneracy for at most four blocks]\ \label{prop:four:blocks}
For $\sigma\in F=\IZ/N\IZ$, written in the form \eqref{sigma:2m} with $2k\le 4$ blocks, one has
\[D_{\rm stab}(\sigma)= D_{\rm sym}(\sigma).\]
\end{prop}
\textbf{Proof:}
$\bullet$ {\bf No blocks:}  $\sigma=\pm \idty$ are the only configurations with 
$\Delta\, A_F(\sigma)=0$.\\
$\bullet$ {\bf Two blocks:}
The configurations \eqref{sigma:2m} with $k=1$, that is  
\[\sigma^{(m)}\in G_F\qmbox{,}
\sigma^{(m)}_\ell=
\left\{
\begin{array}{cll}
1&,&\ell\leq m\\
-1 &,&m<\ell\leq N
\end{array}\right. \qquad(m\in \{1,\ldots,N-1\})\]
have a correlation $A(\sigma^{(m)})_{r}=N\!-\!4\min(\|m\|_1,\|r\|_1)$
of triangular form,
and $\Delta\, A_F(\sigma^{(m)})$ $= \delta_m+\delta_{-m}-2\delta_0$.\\
$D_{\rm stab}(\sigma^{(m)}) = D_{\rm sym}(\sigma^{(m)})$,
since by Remark \ref{rem:reconstuct}.1 for any 
$\tau\in A_F^{-1}\big(A_F(\sigma^{(m)})\big)$
there are $\Delta\, A_F(\tau)_0=\Delta\, A_F(\sigma^{(m)})_0=2$ blocks, which by 
Remark \ref{rem:reconstuct}.3 are of sizes $m$ respectively $N-m$.
So $\tau\in \Phi({\mathcal S},\sigma^{(m)})$.\\[2mm]
$\bullet$ {\bf Four blocks, notation:}
We consider the function $\Delta\, A_F(\sigma):\IZ/N\IZ\to\IZ$ as a signed multiset
\beq 
M:=\prod_{i=1}^{i_{\max}} t_i^{c_i} \qquad \mbox{with } 0<t_i < t_{i+1}
\mbox{ and } c_i\in\IZ\setminus\{0\}.
\Leq{signed:multiset}
By Lemma \ref{lem:Delta:A}, the sum of lengths of all four blocks equals $t_{i_{\max}}=N$, 
and $c_{i_{\max}}=-4$. We denote that
contribution to  \eqref{signed:multiset} by $M_4^{-1}:=N^{-4}$.
The other exponents are palindromic: 
\[ c_{i_{\max}-i}=c_i\quad(i=1,\ldots,i_{\max}-1).\]
$\bullet$ {\bf Four blocks, equality in \eqref{ineq:l1}:}
For $\sigma$ of the form \eqref{sigma:2m} with $k=2$ we start with the assumption
\beq
\|\Delta\, A_F(\sigma)\|_1=(\Delta\, A_F(\sigma))_0^2\equiv 16.
\Leq{ass:k2}

As $M_4^{-1}=N^{-4}$, the contribution to the signed multiset 
$M=M_1M_2^{-1}M_3M_4^{-1}$, coming from pairs of neighbouring blocks, equals
$M_2^{-1}=
\prod_{\stackrel{t=1,\ldots,N-1}{\Delta A_F(\sigma)|_t<0}} t^{\Delta A_F(\sigma)_t}$.

%
%
\begin{table}[htdp]
\caption{Reconstruction from the correlations for four blocks, 
$\|\Delta A_F(\sigma)\|_1=12$}
\medskip

{\scriptsize
\begin{tabular}{| l |c| l |c||c|c|}
\hline
multiset&\!\!\!\!par\!\!\!\!&further conditions&block lengths&\!\!\!\!Example\!\!\!\!&$t_i$ in Example\\
\hline
\hline
$t_1^3t_2^{-1}$ &e& &$t_1,t_1,t_1,t_2$&1,1,1,2&1,2,3,4,5\\
\hline
$t_1^2t_2t_3^{-1}$ &o& &$t_1,t_1,t_2,t_2$&1,1,2,2&1,2,3,4,5\\
\hline
$t_1^2t_2t_3^{-1}$ &e& $ t_3=2t_1,\qquad 3t_1+2t_2=N$&$t_1,t_1,t_2,t_1+t_2$&2,2,3,5&2,3,4,8,9,10\\
$t_1^2t_2t_3^{-1}$ &e& $ t_3=t_1+t_2,\ 2t_1+t_2+t_3=N$ &$t_1,t_2,t_1,t_3$&1,2,1,3&1,2,3,4,5,6\\
$t_1^2t_2t_3^{-1}$ &e& $t_3=3t_1 ,\qquad 4t_1+t_2=N$&$t_1,t_1,2t_1,t_2$&2,2,4,5&2,5,6,7,8,11\\
\hline
$t_1^2t_2^{-1}t_3$ &o& &$t_1,t_1,2t_1,t_3$&1,1,2,4&1,3,4,5,7\\
\hline
$t_1^2t_2^{-1}t_3$ &e& $t_2=2t_1 ,\, 3t_1+2t_3=N$&$t_1,t_1,t_3,t_1+t_3$&1,1,3,4&1,2,3,6,7,8\\[1mm]
$t_1^2t_2^{-1}t_3$ &e&$t_2=3t_1,\,4t_1+t_3=N
$&$t_1,t_1,2t_1,t_3$&2,2,4,7&2,6,7,8,9,13N\\
$t_1^2t_2^{-1}t_3$ &e&$t_2=3t_1,\,4t_1+t_4=N
$ &$t_1,t_1,2t_1,t_4$&1,1,2,5&1,3,4,5,6,8\\
\hline
$t_1t_2^2t_3^{-1}$ &o& &$t_1,t_2,t_2,t_1+t_2$&1,2,2,3&1,2,4,6,7\\
\hline
$t_1t_2^2t_3^{-1}$ &e&$ t_3=t_1+t_2,\ t_1+2t_2+t_3=N$ &$t_1,t_2,t_3,t_2$&1,2,3,2&1,2,3,5,6,7\\
$t_1t_2^2t_3^{-1}$ &e&$ t_3=t_1+t_2,\ t_1+4t_2=N$ &$t_1,t_2,t_2,2t_2$&1,2,2,4&1,2,3,6,7,8\\[1mm]
$t_1t_2^2t_3^{-1}$ &e&$ t_3=2t_1+t_2,\ 2t_1+3t_2=N$ &$t_1,t_2,t_2,t_1+t_2$&1,4,4,5&1,4,6,8,10,13\\
\hline
$t_1t_2t_3t_4^{-1}$ &o& $ t_4=t_1+t_3,\,2t_1+2t_2+t_3=N$&$t_1,t_2,t_1+t_2,t_3$&1,2,3,4&1,2,4,5,6,8,9\\[1mm]
$t_1t_2t_3t_4^{-1}$ &o& $t_4=t_2+t_3,\,2t_1+2t_2+t_3=N$&$t_1,t_2,t_3,t_1+t_2 $&2,3,4,5&2,3,4,7,10,11,12\\
$t_1t_2t_3t_4^{-1}$ &o& $ t_4=t_2+t_3,\,2t_1+t_2+2t_3=N$&$t_1,t_3,t_2,t_1+t_3$&1,3,2,4&1,2,3,5,7,8,9\\
\hline
$t_1t_2t_3t_4^{-1}$ &e& $t_4=t_1+t_2,\, t_1+2t_2+2t_3=N$&$t_1,t_2,t_3,t_2+t_3$&2,3,4,7&2,3,4,5,11,12,13,14\\
$t_1t_2t_3t_4^{-1}$ &e& $t_4=t_1+t_2 ,\, 2t_1+t_2+2t_3=N$&$t_1,t_2,t_1+t_3,t_3$&2,3,6,4&2,3,4,5,7,8,11,12,13\\[1mm]
$t_1t_2t_3t_4^{-1}$ &e& $t_4=t_1+t_3,\, t_1+2t_2+2t_3=N$&$t_1,t_3,t_2,t_2+t_3$&1,3,2,5&1,2,3,4,7,8,9,10\\
$t_1t_2t_3t_4^{-1}$ &e& $t_4=t_1+t_3,\, 2t_1+2t_2+t_3=N$&$t_1,t_2,t_1+t_2,t_3$&1,3,4,5&1,3,5,6,7,8,10,12\\[1mm]
$t_1t_2t_3t_4^{-1}$ &e& $ t_4=t_2+t_3,\, 2t_1+2t_2+t_3=N $&$t_1,t_2,t_3,t_1+t_2$&3,4,5,7&3,4,5,9,10,14,15,16\\[1mm]
$t_1t_2t_3t_4^{-1}$ &e& $t_4=2t_1+t_3 ,\, 2t_1+t_2+2t_3=N$&$t_1,t_3,t_2,t_1+t_3$&1,4,3,5&1,3,4,6,7,9,10,12\\
\hline
$t_1t_2t_3^{-1}t_4$ &o& 
$t_3=2t_1+t_2$&$t_1,t_2,t_4,t_1+t_2$&1,2,6,3&1,2,4,6,8,10,11\\
$t_1t_2t_3^{-1}t_4$ &o&
$t_3=t_1+2t_2$&$t_1,t_2,t_1+t_2,t_4$&1,2,3,6&1,2,5,6,7,10,11\\
\hline
$t_1t_2t_3^{-1}t_4$ &e& $t_3=t_1+t_2,\,\qquad t_1+2t_2+t_3+t_4=N  $&$t_1,t_2,t_2+t_3,t_4$&1,2,5,4&1,2,3,4,8,9,10,11\\
$t_1t_2t_3^{-1}t_4$ &e& $t_3=t_1+t_2,\, \qquad t_1+2t_2+2t_4=N $&$t_1,t_2,t_4,t_2+t_4$&1,2,4,6&1,2,3,4,5,6,7,9,11,12\\
$t_1t_2t_3^{-1}t_4$ &e& $t_3=t_1+t_2,\, t_4=t_2+t_3,\,2t_1+t_2+2t_4=N$&$t_1,t_2,t_1+t_4,t_4$&1,2,6,5&1,2,3,5,9,11,12,13\\[1mm]

$t_1t_2t_3^{-1}t_4$ &e& $t_3=2t_1+t_2,\, 2t_1+2t_2+t_5=N$&$t_1,t_2,t_5,t_1+t_2$&1,2,8,3&1,2,4,6,8,10,12,13\\
$t_1t_2t_3^{-1}t_4$ &e& $t_3=2t_1+t_2,\, t_4=t_1+t_3,\, 
2t_1+2t_2+t_4=N\!\!$&$t_1,t_2,t_4,t_1+t_2$&1,2,5,3&1,2,4,5,6,7,9,10\\[1mm]

$t_1t_2t_3^{-1}t_4$ &e& $t_3=t_1+2t_2,\, 2t_1+2t_2+t_4=N$&$t_1,t_2,t_1+t_2,t_4$&\!\!2,4,6,11\!\!&\!\!2,4,10,11,12,13,19,21\!\!\\
$t_1t_2t_3^{-1}t_4$ &e& $t_3=t_1+2t_2,\, 
2t_1+2t_2+t_5=N$&$t_1,t_2,t_1+t_2,t_5$&\!\!1,2,3,10\!\!&1,2,5,6,10,11,14,15\\

\hline 
\hline
\end{tabular}
\label{table:delta:12}
}
\end{table}

\begin{table}[htdp]
\caption{Reconstruction from the correlations for four blocks,
$\|\Delta A_F(\sigma)\|_1=8$}
\medskip
\begin{center}
{\scriptsize
\begin{tabular}{| l |c|c|c||c|c|}
\hline
multiset&par&further conditions&block lengths&Example&$t_i$ in Example\\
\hline
$t_1^2$&e& &$t_1,t_1,2t_1,3t_1$&1,1,2,3&1,6\\
\hline
$t_1t_2$&e&$t_3=3t_1+3t_2$&$t_1,t_2,t_1+t_2,t_1+2t_2$&1,2,3,5&1,2,9,10\\
$t_1t_2$&e&$t_3=4t_1+2t_2$&$t_1,t_2,2t_1+t_2,t_1+t_2$&1,2,4,3&1,2,8,9\\
\hline
\end{tabular}
\label{table:delta:8}
}
\end{center}
\end{table}
%
%
Conversely, the contribution $M_1M_3$, 
coming from single blocks respectively triples of blocks, equals
$\prod_{\stackrel{t=1,\ldots,N-1}{\Delta A_F(\sigma)|_t>0}} t^{\Delta A_F(\sigma)_t}$.
We write the multiset $M_1M_2M_3$ in the form 
$\tilde{t}_1\cdot\ldots\cdot\tilde{t}_{12}$, with $\tilde{t}_i\le \tilde{t}_{i+1}$ and 
$\tilde{t}_{13-i}=N-\tilde{t}_i$.
Using a permutation
$\rho\in S_4$ so that $m_{\rho(1)}\le m_{\rho(2)}\le m_{\rho(3)}\le m_{\rho(4)}$,
$m_{\rho(1)}=\tilde{t}_1$ and $m_{\rho(2)}=\tilde{t}_2$. 
So in particular $\tilde{t}_1$ and $\tilde{t}_2$ are in $M_1$.
By a suitable choice of $\rho$ in case of degeneracy,
$\rho(1)$ and $\rho(2)$ are neighbouring indices iff $\tilde{t}_1+\tilde{t}_2\in M_2$. 
\begin{itemize}
\item 
Assuming this, $m_{\rho(1)}+m_{\rho(2)}\in \{\tilde{t}_3,\tilde{t}_4,\tilde{t}_5\}$.
If $m_{\rho(1)}+m_{\rho(2)}=\tilde{t}_3$, then $m_{\rho(3)}=\tilde{t}_4$, and otherwise
$m_{\rho(3)}=\tilde{t}_3$. 

If $m_{\rho(i)}+m_{\rho(3)}\in M_2$, then $\rho(i)$ and $\rho(3)$ are 
neighbouring indices ($i=1,2$), and by assumption \eqref{ass:k2} 
exactly one of these alternatives is true.

In both cases, $m_{\rho(4)}=N-m_{\rho(1)}-m_{\rho(2)}-m_{\rho(3)}$.
\item 
Assuming instead that $m_{\rho(1)}+m_{\rho(2)}\not\in M_2$, 
$m_{\rho(3)}=\tilde{t}_3$. Then $M_2$ contains $\tilde{t}_1+\tilde{t}_3$ and 
$\tilde{t}_2+\tilde{t}_3$, the remaining
two elements of $M_2$ being of the form $\tilde{t}_1+m_{\rho(4)}\le \tilde{t}_2+m_{\rho(4)}$.
By subtracting $\tilde{t}_1$ from $\tilde{t}_1+m_{\rho(4)}$, we get $m_{\rho(4)}$.
\end{itemize}
In any case we identified the sequence $(m_1,m_2,m_3,m_4)$ of block lengths, up to
the action of the dihedral subgroup $D_4$ of $S_4$.\\[2mm]
$\bullet$ {\bf Four blocks, strict inequality in \eqref{ineq:l1}:}
By Lemma \ref{lem:l1:norm} we are left to consider the values 12 and 8 of 
$\|\Delta\, A_F(\sigma)\|_1$.  The forms of signed multisets  leading to these values
are listed in the first column of Table \ref{table:delta:12} resp.\ \ref{table:delta:8}. 
Because their exponents are palindromic,
we list only the first half, and (in Column 2)
the information whether $i_{\max}$ is even (parity e) or odd (parity o).

In many cases knowledge of the exponents $c_i$ does not suffice 
to reconstruct the block lengths up to symmetry and thus to
show equality of stable and symmetry-induced degeneracy. 
In these cases the mutually exclusive further conditions in Column 3 allow
decoding of the sequence $(m_1,m_2,m_3,m_4)$ of block lengths (Column 4). 
Finally Column 5 gives examples of realizations $(m_1,m_2,m_3,m_4)$, and 
their signed multisets (Column 6).
\hfill $\Box$\\[2mm]

A more conceptual proof would be welcome, since it could help to find
general upper bounds for stable degeneracies and to verify Conjecture \eqref{conjecture}.

For some $\sigma\in G_F$ with $F=\IZ/N\IZ$ their $\Phi$--orbit (see \eqref{Phi})
can be reconstructed from the correlation $A_F(\sigma)$ (so that in particular
$D_{\rm stab}(\sigma) = D_{\rm sym}(\sigma)$). 
There is an analog of the main theorem of Ginzburg and Rudnick in \cite{GR} valid
for Ising chains.
\begin{prop}[Reconstruction from the correlation]\label{prop:reconstruction}
Assume that for $\sigma$ of the form \eqref{sigma:2m}
the map \ ${\cal P}(\{1,\ldots,2k\})\to \IN,\ I\mapsto \sum_{i\in I}m_i$ \ is injective.\\
Then given $\Delta A_F(\sigma)$, the orbit $\Phi({\cal S},\sigma)$ of $\sigma$ can be calculated.
\end{prop}
\textbf{Proof:}
$\bullet$
We can assume w.l.o.g.\ that $\sigma \neq \pm\idty_F$, that is, $k>0$.\\
$\bullet$
If $\Delta A_F(\sigma')=\Delta A_F(\sigma)$ for a $\sigma'\in G_F$, then 
by Remark \ref{rem:reconstuct}.1 the number of blocks of $\sigma'$ equals $2k$, too.
We assume (by applying the $\Phi$ action, if necessary) that $\sigma'$ 
is of the form \eqref{sigma:2m}, that is
\[ \sigma' = (1)^{m'_1}(-1)^{m'_2}\ldots(1)^{m'_{2k-1}}(-1)^{m'_{2k}}.\]
$\bullet$
The next task is to show that $(m'_1,\ldots, m'_{2k})$ is a permutation of
$(m_1,\ldots, m_{2k})$. We identify $(\ell,n)\in X:=\{1,\ldots,2k\}\times \{1,\ldots,2k-1\}$
with the subinterval $\{\ell,\ldots,\ell+n-1\}\subsetneq \IZ/(2k\IZ)$.
Then there is a unique bijection 
\beq
B: X\to X\qmbox{with} \sum_{i\in \{\ell,\ldots,\ell+n-1\}}
\!\!\!\!\!m_i = \sum_{i\in B(\{\ell,\ldots,\ell+n-1\})}\!\!\!\!\! m'_i\qquad
\big((\ell,n)\in X\big) .
\Leq{def:B}
Let for $\alpha,\beta\in S_{2k}$ the sequences
$(m_{\alpha(1)},\ldots, m_{\alpha(2k)})$ and $(m'_{\beta(1)},\ldots, m'_{\beta(2k)})$
be weakly ascending. By the assumption of the proposition then
$(m_{\alpha(1)},\ldots, m_{\alpha(2k)})$ and thus also $(m'_{\beta(1)},\ldots, m'_{\beta(2k)})$
are strictly ascending. Then $m_{\alpha(1)}=m'_{\beta(1)}$. Contradicting our hypothesis
that the sets $\{m'_1,\ldots, m'_{2k}\}$ and $\{m_1,\ldots, m_{2k}\}$ coincide, let
now $\ell$ be the first integer so that $m_{\alpha(\ell)}\neq m'_{\beta(\ell)}$.
Then $m_{\alpha(\ell)}$ is the sum of at least two integers $m'_i$ whose indices $i$ are 
of the form $\beta(j)$ with $j<\ell$. So $m_{\alpha(\ell)}=\sum_{i\in I} m_{\alpha(i)}$ with
index set $I\subseteq \{1,\ldots,\ell-1\}$. But this contradicts the assumption
of the proposition, proving $ m'_{\beta(\ell)} = m_{\alpha(\ell)}$
or $m'_{\gamma (\ell)}=m_\ell\quad(\ell=1,\ldots,2k)$ 
with $\gamma:=\beta\circ\alpha^{-1}\in S_{2k}$.\\[2mm]
$\bullet$
Finally we show that $\sigma'\in  \Phi({\cal S},\sigma)$.
By again applying the $\Phi$ action, if necessary, we can assume that $\gamma(1)=1$,
so that $m'_1=m_1$. For all $i\in \{1,\ldots,2k\}$ the interval
$\{i,i+1\}$ is mapped by $B$ onto an interval which is also of length two, and which is moreover
of the form $\{\gamma(i),\gamma(i+1)\}$, since otherwise \eqref{def:B} would contradict
the assumption of the proposition.

However this means that $\gamma$ belongs to the dihedral subgroup $D_{2k}$ of
the permutation group $S_{2k}$.
\hfill $\Box$


\begin{thebibliography}{99}
\addcontentsline{toc}{section}{References}

\bibitem[Be]{Be}
Beale, Paul:
Exact distribution of energies in the two-dimensional Ising model.
Physical Review Letters  {\bf 76}, 78--81 (1996)

\bibitem[BS]{BS}
Bach, Eric; Shallit, Jeffrey:
Algorithmic Number Theory: Efficient algorithms, Vol. 1.
The MIT Press, 1997

\bibitem[BZ]{BZ}
Baier, Stephan; Zhao, Liangyi:
On primes represented by quadratic polynomials.
Proceedings of the 'Anatomy of Integers' Conference (Montreal, 2006), 159--166 (2007)

\bibitem[Ca]{Ca}
C\u{a}lug\u{a}reanu, Grigore:
The total number of subgroups of a finite Abelian group.
Scientiae Mathematicae Japonicae {\bf 10}, 207--217 (2003)

\bibitem[GR]{GR}
Ginzburg, Viktor; Rudnick, Zeev:
Stable multiplicities in the length spectrum of Riemann surfaces.
Israel Journal of Mathematics {\bf 104}, 129--144 (1998)

\bibitem[Go]{Go}
Golay, Marcel: 
Notes on the Representation of $1 , 2 ,\ldots, n$ by Differences.
J. London Math. Society {\bf 4}, 729--734 (1972)

\bibitem[HL]{HL}
Halberstam, Heine; Laxton, Robert:
On Perfect Difference Sets.
The Quarterly Journal of Mathematics {\bf 14}, 86--90 (1963)

\bibitem[Ho]{Ho}
Horowitz, Robert: 
Characters of free groups represented in the two-dimensional special linear group.
Comm. Pure Appl. Math. {\bf 25}, 635--649 (1972)

\bibitem[HP]{HP}
Hughes, Daniel; Piper, Frederick:
Projective Planes (Graduate Texts in Mathematics {\bf 6}) 
New York,   Springer-Verlag, 1982 

\bibitem[JJ]{JJ}
Jacobs, Konrad; Jungnickel, Dieter: 
Einf\"uhrung in die Kombinatorik. Berlin, De Gruyter, 2004

\bibitem[Ka]{Ka}
Kaufman, Bruria:
Crystal statistics. II. Partition function evaluated by spinor analysis.
Physical Review {\bf 76}, 1232--1243 (1949)

\bibitem[LV]{LV}
Loebl, Martin; Vondr\'ak, Jan:
Towards a theory of frustrated degeneracy.
Discrete Mathematics {\bf 271}, 179--193 (2003) 

\bibitem[Ra]{Ra}
Randol, Burton:
The Length Spectrum of a Riemann Surface is Always of Unbounded Multiplicity.
Proceedings of the AMS {\bf 78}, 455--456 (1980)

\bibitem[Si]{Si}
Singer, James:
A Theorem in Finite Projective Geometry and some Applications to Number Theory.
Transactions of the American Mathematical Society {\bf 43}, 377--385 (1938)
\end{thebibliography}
\end{document}